\documentclass[twoside,twocolumn,9pt]{article}
\usepackage{extsizes}
\usepackage[super,sort&compress,comma]{natbib} 
\usepackage[version=3]{mhchem}
\usepackage[left=1.5cm, right=1.5cm, top=1.785cm, bottom=2.0cm]{geometry}
\usepackage{balance}
\usepackage{times,mathptmx}
\usepackage{sectsty}
\usepackage{graphicx} 
\usepackage{lastpage}
\usepackage[format=plain,justification=justified,singlelinecheck=false,font={stretch=1.125,small,sf},labelfont=bf,labelsep=space]{caption}
\usepackage{float}
\usepackage{fancyhdr}
\usepackage{fnpos}
\usepackage[english]{babel}
\addto{\captionsenglish}{%
  
}
\usepackage{array}
\usepackage{droidsans}
\usepackage{charter}
\usepackage[T1]{fontenc}
\usepackage[usenames,dvipsnames]{xcolor}
\usepackage{setspace}
\usepackage[compact]{titlesec}
\usepackage{hyperref}
\usepackage{epstopdf}

\usepackage{graphicx}
\usepackage{dcolumn}
\usepackage{bm}
\usepackage[utf8]{inputenc}
\usepackage{graphicx}
\usepackage{color}
\usepackage{amsmath}
\usepackage{amssymb}
\usepackage{subfigure}
\usepackage{epsfig}
\usepackage{float}
\usepackage{lipsum}
\usepackage{cases}
\usepackage{appendix}
\usepackage{cancel}
\usepackage{soul}

\newcommand{\jj}{\vec{j}}

\newcommand{\sR}{\vec{\mathcal{R}}}

\newcommand{\Gama}{{\mathpalette\makeGama\relax}}
\newcommand{\makeGama}[2]{%
  \raisebox{\depth}{\scalebox{1}[-1]{$\mathsurround=0pt#1\mathbb{L}$}}%
}

\newcommand{\M}{\mathrm{\mathbb{M}}}
\newcommand{\sigg}{\vec{\sigma}}

\newcommand{\B}{\vec{B}}

\newcommand{\1}{\mathbb{I}}

\newcommand{\J}{\vec{J}}

\newcommand{\dif}{\mathop{}\!\mathrm{d}}

\newcommand{\K}{\mathop{}\!\mathrm{K}}

\newcommand{\sign}{\mathop{}\!\boldsymbol{\mathrm{sign}}}

\newcommand{\p}{\vec{p}}
\newcommand{\pii}{\vec{\pi}}
\newcommand{\vv}{\vec{v}} 

\newcommand{\rr}{\vec{r}}

\newcommand{\xxi}{\vec{\xi}}

\newcommand{\eeta}{\vec{\eta}}

\definecolor{cream}{RGB}{222,217,201}

\begin{document}

\pagestyle{fancy}
\thispagestyle{plain}
\fancypagestyle{plain}{

\renewcommand{\headrulewidth}{0pt}
}

\makeFNbottom
\makeatletter
\renewcommand\LARGE{\@setfontsize\LARGE{15pt}{17}}
\renewcommand\Large{\@setfontsize\Large{12pt}{14}}
\renewcommand\large{\@setfontsize\large{10pt}{12}}
\renewcommand\footnotesize{\@setfontsize\footnotesize{7pt}{10}}
\makeatother

\renewcommand{\thefootnote}{\fnsymbol{footnote}}
\renewcommand\footnoterule{\vspace*{1pt}%
\color{cream}\hrule width 3.5in height 0.4pt \color{black}\vspace*{5pt}} 
\setcounter{secnumdepth}{5}

\makeatletter 
\renewcommand\@biblabel[1]{#1}            
\renewcommand\@makefntext[1]%
{\noindent\makebox[0pt][r]{\@thefnmark\,}#1}
\makeatother 
\renewcommand{\figurename}{\small{Fig.}~}
\sectionfont{\sffamily\Large}
\subsectionfont{\normalsize}
\subsubsectionfont{\bf}
\setstretch{1.125} 
\setlength{\skip\footins}{0.8cm}
\setlength{\footnotesep}{0.25cm}
\setlength{\jot}{10pt}
\titlespacing*{\section}{0pt}{4pt}{4pt}
\titlespacing*{\subsection}{0pt}{15pt}{1pt}

\fancyfoot{}
\fancyfoot[RO]{\footnotesize{\sffamily{1--\pageref{LastPage} ~\textbar  \hspace{2pt}\thepage}}}
\fancyfoot[LE]{\footnotesize{\sffamily{\thepage~\textbar\hspace{2pt} 1--\pageref{LastPage}}}}
\fancyhead{}
\renewcommand{\headrulewidth}{0pt} 
\renewcommand{\footrulewidth}{0pt}
\setlength{\arrayrulewidth}{1pt}
\setlength{\columnsep}{6.5mm}
\setlength\bibsep{1pt}

\makeatletter 
\newlength{\figrulesep} 
\setlength{\figrulesep}{0.5\textfloatsep} 

\newcommand{\topfigrule}{\vspace*{-1pt}%
\noindent{\color{cream}\rule[-\figrulesep]{\columnwidth}{1.5pt}} }

\newcommand{\botfigrule}{\vspace*{-2pt}%
\noindent{\color{cream}\rule[\figrulesep]{\columnwidth}{1.5pt}} }

\newcommand{\dblfigrule}{\vspace*{-1pt}%
\noindent{\color{cream}\rule[-\figrulesep]{\textwidth}{1.5pt}} }

\makeatother

\twocolumn[
  \begin{@twocolumnfalse}
\vspace{3cm}
\sffamily
\begin{tabular}{m{4.5cm} p{13.5cm} }

& \noindent\LARGE{\textbf{Stochastic resetting of active Brownian particles with Lorentz force}} \\
\vspace{0.3cm} & \vspace{0.3cm} \\

 & \noindent\large{Iman Abdoli\textit{$^{a}$} and Abhinav Sharma$^{\ast}$\textit{$^{ab}$} } \\

 & \noindent\normalsize{The equilibrium properties of a system of passive diffusing particles in an external magnetic field are unaffected by the Lorentz force.
In contrast, active Brownian particles
exhibit steady-state phenomena that depend on both the strength and the polarity of the applied magnetic field. The intriguing effects of the Lorentz force, however, can only be observed when out-of-equilibrium density gradients are maintained in the system. To this end, we use the method of stochastic resetting on active Brownian particles in two dimensions by resetting them to the line $x=0$ at a constant rate and periodicity in the $y$ direction. Under stochastic resetting, an active system settles into a nontrivial stationary state which is characterized by an inhomogeneous density distribution, polarization and bulk fluxes perpendicular to the density gradients. We show that whereas for a uniform magnetic field the properties of the stationary state of the active system can be obtained from its passive counterpart, novel features emerge in the case of an inhomogeneous magnetic field which have no counterpart in passive systems. In particular, there exists an activity-dependent threshold rate such that for smaller resetting rates, the density distribution of active particles becomes non-monotonic. We also study the mean first-passage time to the $x$ axis and find a surprising result: it takes an active particle more time to reach the target from any given point for the case when the magnetic field increases away from the axis. The theoretical predictions are validated using Brownian dynamics simulations.

} \\

\end{tabular}

 \end{@twocolumnfalse} \vspace{0.6cm}

  ]

\renewcommand*\rmdefault{bch}\normalfont\upshape
\rmfamily
\section*{}
\vspace{-1cm}

\footnotetext{\textit{$^{a}$~Leibniz-Institut  f\"ur Polymerforschung Dresden, Institut Theorie der Polymere, 01069 Dresden, Germany; E-mail: sharma@ipfdd.de}}
\footnotetext{\textit{$^{b}$~Technische Universit\"at Dresden, Institut f\"ur Theoretische Physik, 01069 Dresden, Germany}}



\section{Introduction}
\label{interoduction} 
A fundamental feature of Active Brownian Particles (ABPs) is self-propulsion which requires a continual consumption of energy from the local environment ~\cite{fiasconaro2008active,romanczuk2012active, wensink2013differently,walther2013janus, wang2013nanomachines,volpe2014simulation, elgeti2015physics}. Since ABPs are internally driven, they do not require breaking the spatial symmetry to exist in an out-of-equilibrium state. An ABP is generally modelled as a particle which propels itself along a direction which randomizes via rotational diffusion. Given the simplicity of the model, it is not surprising that ABPs serve as a minimalistic model to study the effect of broken time-reversal symmetry and nonequilibrium steady states in general~\cite{cates2012diffusive,fodor2016far,cates2015motility,digregorio2018full, mandal2019motility,lowen2020inertial, singh2020phase}. The interest in ABPs is not purely theoretical as evident in the vast body of research in pharmaceutical and medical applications~\cite{sezer2016smart, yang2012janus,paxton2004catalytic,fournier2005synthetic, howse2007self,ebbens2010pursuit, poon2013clarkia,ekeh2020thermodynamic}. Since an ABP adjusts its propulsion speed in response to the local fuel concentration~\cite{howse2007self, gao2014catalytic}, it is also used as a simple model to understand the emergence of chemotaxis in proto-forms of life~\cite{ghosh2015pseudochemotactic, vuijk2018pseudochemotaxis,merlitz2020pseudo,vuijk2020chemotaxis}.

Recently, the behaviour of diffusion systems subjected to an external magnetic field has attracted considerable interest~\cite{filliger2007kramers, balakrishnan2008elements, friz2015physical, chun2018emergence, vuijk2019effect,vuijk2019anomalous, chun2019effect, abdoli2020nondiffusive, vuijk2019lorenz,abdoli2020stationary, abdoli2020correlations, cerrai2020averaging}. It has been shown that Lorentz force due to an external magnetic field induces additional Lorentz fluxes in diffusion systems which are perpendicular to the typical diffusive fluxes~\cite{ vuijk2019anomalous, abdoli2020nondiffusive}. The Lorentz force generates dynamics which are different from those of a purely diffusive system. Interestingly, the unusual properties due to the Lorentz force persist in the small-mass limit in which the dynamics are overdamped. However, the equilibrium properties, as expected from the Bohr-van Leeuwen theorem~\citep{van1921problemes}, are unaffected by the applied magnetic field due to no performance of work on the particle. Since the Lorentz force only influences the dynamics, there are essentially two conditions to observe its unusual effects: (i) the system is out of equilibrium and (ii) there are density gradients in the system. This has been recently demonstrated in a system of ABPs with a uniform activity subjected to an inhomogeneous magnetic field~\cite{vuijk2019lorenz}, which satisfies the aforementioned conditions even in the stationary state. The nonequilibrium steady state in such a system is characterized by density inhomogeneity and bulk fluxes.

In order to ensure that there exists a nontrivial stationary state in a system of ABPs, one requires either a confining potential or periodic boundary conditions~\cite{vuijk2019lorenz}. In recent years, stochastic resetting has emerged as a powerful framework which gives rise to nontrivial stationary states in diffusive systems characterized by a non-Gaussian probability distribution and steady-state fluxes~\cite{evans2011diffusion,evans2011optimal,evans2013optimal, durang2014statistical, majumdar2015dynamical, evans2018run,masoliver2019telegraphic, pal2019time,da2019diffusions, evans2020stochastic,magoni2020ising,gupta2020work,belan2020median}. Stochastic resetting is unique in the sense that it renews the underlying process and therefore, in some sense, preserves the dynamics of the underlying process in the steady state. With the recent experimental demonstrations~\cite{tal2020experimental,PhysRevResearch.2.032029}, stochastic resetting is now no longer a pure theoretical pursuit but rather an alternative and practical method to drive and maintain a system out of equilibrium. 
We have recently shown that stochastically resetting a passive particle to the origin in the presence of Lorentz force gives rise to a novel stationary state which bears the unusual dynamical properties owing to the magnetic field~\cite{abdoli2020stationary}. While the stochastic resetting of passive particles has been thoroughly studied, much less is done about active particles~\cite{evans2018run,scacchi2018mean,santra2020run}. In a very recent work, the motion of an ABP under different resetting protocols has been studied~\cite{kumar2020active} with a focus on the steady-state density distribution.

In the present work, we investigate the motion of a charged ABP under resetting and the effect of Lorentz force. The particle is stochastically reset to the line $x=0$ at a constant rate. In addition, the system is periodic in the $y$ direction. We start with a generalized coarse-grained Fokker-Planck equation and analytically determine the density, flux, and polarization, first for a uniform magnetic field and then for a spatially inhomogeneous magnetic field. We show that whereas for a uniform magnetic field the properties of the stationary state of the active system can be obtained from its passive counterpart, novel features emerge in the case of an inhomogeneous magnetic field which have no counterpart in passive systems. In particular, there exists an activity-dependent threshold rate such that for smaller resetting rates, the density distribution of active particles becomes non-monotonic. We also study the Mean First-Passage Time (MFPT) to the $x$ axis and find a surprising result: it takes an active particle more time to reach the target from any given point for the case when magnetic field increases away from the axis.

The paper continues as follow. In Sec.~\ref{Model}, we define the model and provide a description of the methods used to analyze the system. In Sec.~\ref{uniform}, we study the system in the presence of a constant magnetic field. We then consider a space-dependent magnetic field and derive expressions for the density, flux, and polarization in the system in Sec.~\ref{spacedependent}. In Sec.~\ref{MFPT_sec}, we obtain the MFPT for the active and passive systems. The conclusion of the paper is presented in Sec.~\ref{conclusion}.

\section{Model and theory}
\label{Model}
We consider a single self-propelled, charged Brownian particle of mass $m$ and charge $q$ subjected to an external magnetic field $\B(\rr)$ of strength $B(\rr)$, whose direction is along the $z$ axis where the Lorentz force does not influence the motion of the particle. As a consequence, we study the dynamics of the particle in the $xy$ plane with $\rr=(x, y)$. The particle is stochastically reset to the line $x=0$ at a constant rate $\mu$. The generalized Fokker-Planck equation for the probability density of finding the particle at position $\rr$ with orientation $\p=(p_x, p_y)$ at time $t$ given that the particle started its motion at the origin, $P(\rr, \p;t)$ is given as
\begin{align}
\label{fullfpe}
\frac{\partial}{\partial t} P(\rr, \p;t) &  =  \nabla\cdot\left[\Gama^{-1}(\rr)\cdot \left(D_t\nabla-v_0\p\right)P(\rr, \p;t)\right] \nonumber \\ 
 & + D_r \sR^2 P(\rr, \p;t) + \Phi_l + \Phi_g,  
\end{align}
where $\nabla=(\partial_x, \partial_y)$ and 
\begin{equation}
\label{loss}
\Phi_l = - \mu P(\rr, \p;t),
\end{equation}
is the loss of the probability from the position $\rr$ due to resetting while
\begin{equation}
\label{gain}
\Phi_g = \mu\delta(x)\int P(x',y, p_x,p_y;t)\dif x',
\end{equation}
is the gain of the probability at the point $(0,y)$ on the $x$-axis. Although Eq.~\eqref{fullfpe} is not of the form of a continuity equation, the total probability is conserved. Here  $v_0=f/\gamma$ is the self-propulsion speed where $\gamma$ is the friction coefficient and $f$ is the magnitude of the self-propulsion force that drives the particle into the direction of its (unit) orientation vector $\p$. In addition, $\sR=\p\times\nabla_{\p}$ is the rotation operator, $D_t=k_BT/\gamma$ with $k_B$ being the Boltzmann constant is the translational diffusion coefficient and $D_r$ is the rotational diffusion coefficient. The matrix $\Gama$ is defined as $\1+\kappa(\rr)\M$ where $\1$ is the identity matrix, the dimensionless parameter $\kappa(\rr)=qB(\rr)/\gamma$ quantifies the strength of the Lorentz force relative to the frictional force and $\M$ is a matrix with elements $M_{ij}=-\epsilon_{ijk}n_k$ where $\epsilon_{ijk}$ is the antisymmetric Levi-Civita symbol in two dimensions and $n_k$ is the $k$ component of the unit vector $\boldsymbol{n}$ along which the magnetic field is pointed. The inverse of $\Gama$ reads 
\begin{equation}
\label{gammainverse}
\Gama^{-1}(\rr) = \1-\frac{\kappa(\rr)}{1+\kappa^2(\rr)}\M + \frac{\kappa(\rr)^2}{1+\kappa^2(\rr)}\M^2.
\end{equation}
Note that the orientation of the particle remains unchanged under resetting; the particle restarts its motion with the orientation that it had at the time of resetting.

\begin{figure}
\centering
\resizebox*{1\linewidth}{5.5cm}{\includegraphics{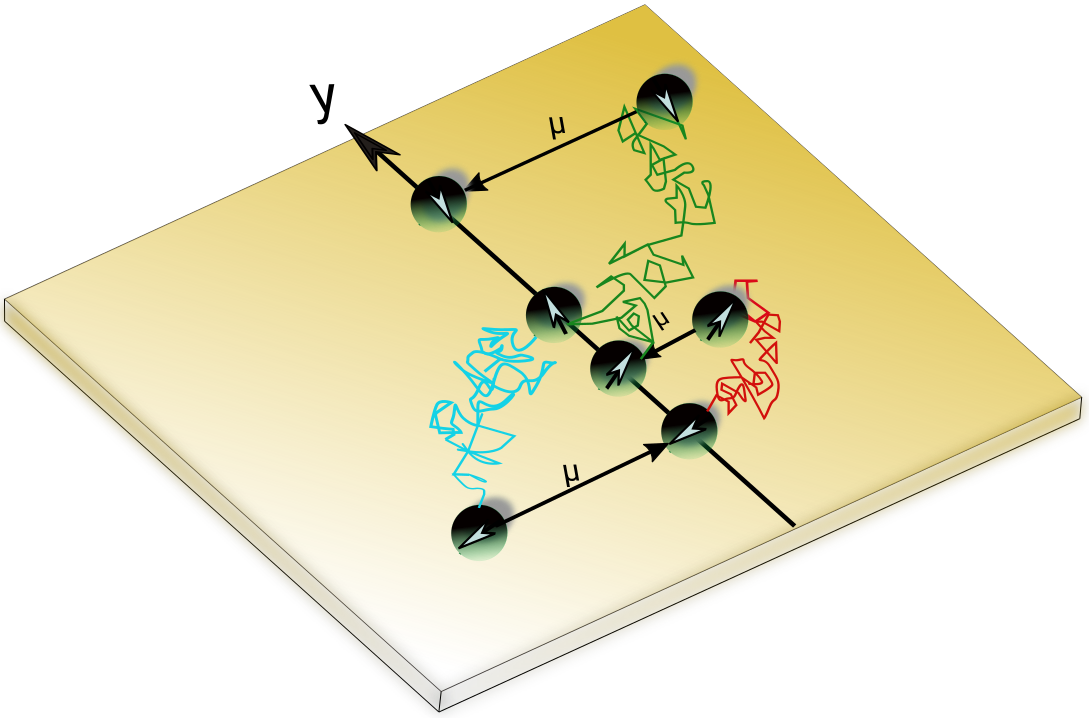}}
\caption{Schematic of a charged active Brownian particle which is stochastically reset to the line $x=0$ at a constant rate $\mu$. The self-propulsion velocity is shown by an arrow inside the disc. Between any two consecutive resetting events the particle undergoes Brownian motion and self-propulsion. Immediately after a resetting event, the orientation of the ABP remains unchanged. The system is subjected to an external magnetic field, $B(\rr)$ in the $z$ direction. } 
\label{schema}
\end{figure}
We also perform Brownian dynamics simulations to validate our theoretical predictions. The dynamics of the particle can be described by the following Langevin equations
\begin{align}
\label{langevinv}
\frac{m}{\gamma}\dot \vv(t) &= - \Gama(\rr)\cdot\vv + v_0 \p(t) + \xxi(t), \\ \label{Langevinp}
\dot \rr(t) &= \vv(t), \,\,\, \text{and}\,\,\, \dot\p(t) = \eeta(t)\times\p(t),
\end{align}
where the dot over the vectors denotes the time derivative and the stochastic forces $\xxi(t)$ and $\eeta(t)$ satisfy the properties of Gaussian white noise with zero mean value and correlation functions $\langle\xxi(t)\xxi^{T}(t')\rangle=2D_t\1\delta(t-t')$ and $\langle\eeta(t)\eeta^{T}(t')\rangle=2D_r\1\delta(t-t')$. As the resetting mechanism we consider Poisson distribution for the resetting time which gives the probability of the number of resets to the line $x=0$ in a small interval of time with a constant rate $\mu$ (see Fig.~\ref{schema}). We numerically integrate the set of equations in \eqref{langevinv} and \eqref{Langevinp} with a small mass $m=0.002$ and the integration time step  $dt = 10^{-6}\tau$ where $\tau=\gamma/k_BT$ is the time the particle
takes to diffuse over a unit distance. We also fix $k_B=\gamma=1.0$, the self-propulsion force $f=10.0$, and $D_r=20.0$. The particle starts its motion at the origin with the initial velocity $(v_{0x}, v_{0y})=(1.0,1.0)$ and initial orientation $(p_{0x}, p_{0y})=(1.0, 0.0)$. The choice of the parameters holds throughout the paper.

The Fokker-Planck equation in \eqref{fullfpe} provides a full statistical description of the position and orientation of an ABP under stochastic resetting. However, it is formidable task to obtain an exact solution of this equation. To theoretically analyze the system we make the following assumptions:  (1) $D_r \gg \mu$ and (2) the gradients in the system are small on the length scale of persistence length of the ABP. Under these assumptions, one can integrate out the orientational degrees of freedom by a gradient expansion (see Appendix~\ref{appendixA} for details) to yield an equation for the (marginal) probability density as a function of time and position degrees of freedom alone~\cite{vuijk2019lorenz}. 
The final equation for the coarse-grained density reads as

\begin{align}
\label{fpe}
\frac{\partial \rho(\rr;t)}{\partial t}  = \nabla\cdot\left[\Gama^{-1}(\rr)\cdot\left(D_t\nabla+v_0\pii(\rr;t)\right)\rho(\rr;t)\right] +\phi_l +\phi_g,
\end{align}
where 
\begin{equation}
\label{loss}
\phi_l = - \mu \rho(\rr;t),
\end{equation}
\begin{equation}
\label{gain}
\phi_g = \mu\delta(x)\int \rho(x',y;t)\dif x',
\end{equation}
are the loss and gain of probabilities and $\rho(\rr;t)$ is the (marginal) probability density of finding the particle at position $\rr$ at time $t$ given that the particle started its motion at the origin. The first and second terms in the square brackets are the (negative) probability fluxes stemming from the thermal fluctuations and activity, respectively. The polarization, $\pii(\rr;t)$, defined as the average orientation per particle, is given as 

\begin{equation}
\label{polarization}
\pii(\rr;t) = -\frac{l_p}{2\rho(\rr;t)}\nabla\cdot[\Gama^{-1}(\rr)\rho(\rr;t)],
\end{equation}
where $l_p=v_0/(D_r+\mu)$ denotes the modified persistence length of the ABP. An alternative approach to the above derivation is to treat activity as a perturbation and use the linear response theory for ABPs as outlined in Refs.~\cite{sharma2016communication,merlitz2018linear}. The gradient expansion approach, in contrast, does not require the activity to be small but only that the gradients be small compared to the persistence length of the ABP. It therefore allows one to consider an active system in which the activity dominates over thermal fluctuations.

Now we consider periodic boundary conditions in the $y$ direction and a magnetic field which is varying along the $x$ direction. With these choices we effectively restrict ourselves to spatially single-variable analysis.
By averaging over the $y$ positional degrees of freedom from the Fokker-Planck equation in \eqref{fpe}, we obtain 
\begin{equation}
\label{fpeoned}
\frac{\partial g(x;t)}{\partial t} = -\nabla\cdot\left[\jj(x;t)+\jj^{a}(x;t)\right]-\mu g(x;t)+\frac{\mu}{L}\delta(x),
\end{equation}
where $L$ is the size of the system and $g(x;t)$ is the probability density of finding the particle at position $x$ at time $t$ given that its initial position was at $x=0$. The flux due to thermal fluctuations is
\begin{subequations} \label{fluxes1d}
\begin{equation}
    \jj(x;t) = - D_t\Gama^{-1}(x) \nabla g(x;t),
     \label{fluxth1d}
\end{equation}
and

\begin{equation}
    \jj^{a}(x;t) = - v_0\Gama^{-1}(x) \p(x;t) g(x;t).
     \label{fluxa1d}
\end{equation}
\end{subequations} 
is the flux due to activity where $\nabla g(x;t)=(\partial_xg(x;t), 0)^\top$, and 
\begin{equation}
\label{polarization}
\p(x;t) = -\frac{l_p}{2g(x;t)}\nabla\cdot[\Gama^{-1}(x)g(x;t)],
\end{equation}
is the polarization. Note that since there is no variation in the $y$ direction all the derivatives with respect to $y$ are zero resulting in the reduction of $\nabla$ in Eqs.\eqref{fpeoned} and \eqref{polarization} to simply the derivative with respect to $x$.  


To highlight the new features which emerge in the system of ABPs we make a comparison between the active system and its passive counterpart. As the active system we consider the motion of the particle purely due to the activity by ignoring the thermal term (i.e., $D_t=0$) in Eqs.~\eqref{langevinv} and ~\eqref{Langevinp} and ~\eqref{fluxth1d}. We compare the active system with the passive one wherein the motion of the particle is due to the thermal fluctuations. The governing Langevin equations and corresponding Fokker-Planck equation of the passive system can be obtained by setting the self-propulsion velocity, $v_0$ to zero in Eqs.~\eqref{langevinv} and ~\eqref{Langevinp} and ~\eqref{fluxa1d}.

\section{Uniform magnetic field}
\label{uniform}

We first consider the system subjected to a uniform magnetic field $\kappa(x)\equiv\kappa$. For the active system the stationary probability density, denoted by $g^a(x)$, can be easily obtained by plugging Eq.~\eqref{fluxa1d} into Eq.~\eqref{fpeoned} and setting $\partial_t g(x;t)=0$. The solution can be written as
\begin{equation}
\label{activeuniform}
g^a(x) = \frac{\alpha}{2L}\exp\left(-\alpha|x|\right),
\end{equation}  
where $\alpha=\sqrt{1+\kappa^2}\alpha_a$ with $\alpha_a=\sqrt{\mu/D_a}$ and $D_a=v_0^2/2(D_r+\mu)$ being the modified active diffusivity. The stationary solution in \eqref{activeuniform} is the same as that of the passive system wherein $D_a$ is replaced by $D_t$~\cite{abdoli2020stationary}. 

The polarization can be obtained by plugging Eq.~\eqref{activeuniform} in Eq.~\eqref{polarization}, which in the $x$ direction can be written as
\begin{subequations} \label{polarizationuniform}
\begin{equation}
    p_x(x) = \frac{l_p\alpha_a}{2(1+\kappa^2)}\sign(x),
     \label{pxuniform}
\end{equation}
and in spite of the translational invariance in the $y$ direction there exists polarization which is given as 
\begin{equation}
    p_y(x) = \kappa p_x(x),
     \label{pyuniform}
\end{equation}
\end{subequations} 
However, the substitution of the polarization into Eq.~\eqref{fluxa1d} gives zero fluxes in $y$ direction and 

\begin{equation}
\label{activeuniformflux}
j_x^a(x) = \frac{\mu}{2L}\sign(x)\exp\left(-\alpha|x|\right),
\end{equation}  
in the $x$ direction where $\sign(.)$ denotes the sign function. Note that the stationary polarization and fluxes in the $y$ direction are zero in the absence of the magnetic field.

\begin{figure}[t]
\centering
\resizebox*{1\linewidth}{5cm}{\includegraphics{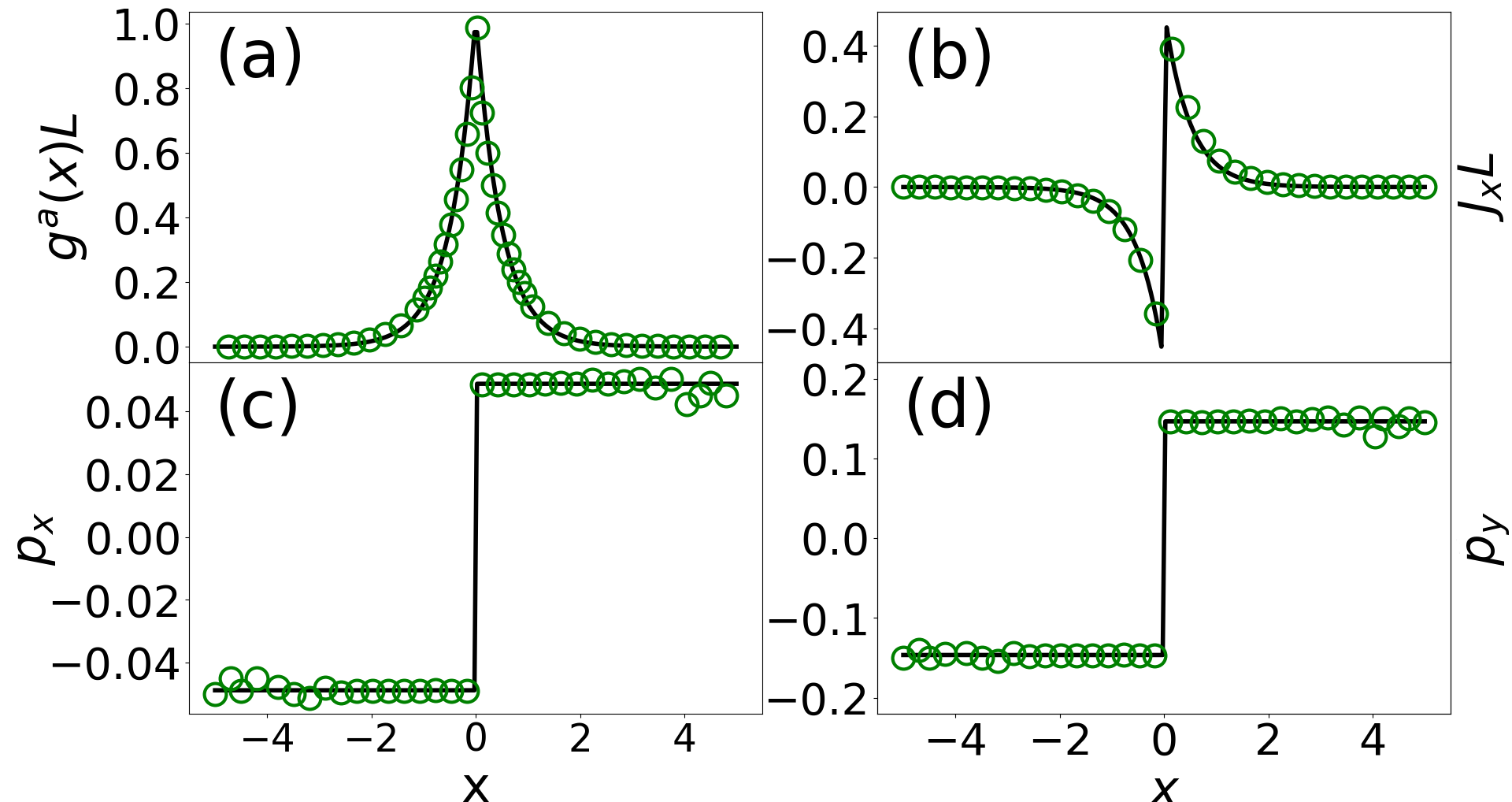}}
\caption{Density, flux in the $x$ direction, and orientations in the $x$ and $y$ directions are shown in (a) to (d), respectively. An ABP is subjected to a constant magnetic field such that $\kappa=3.0$ and is stochastically reset to the line $x=0$ at the rate $\mu=1.0$. Despite the translational invariant motion in the $y$ direction the magnetic field induces polarization in the $y$ direction. However, the $y$ component of the stationary flux is zero due to the cancellation of fluxes arising from the polarization in the $x$ and $y$ directions. The solid lines show the analytical solutions from Eqs.~\eqref{activeuniform} to \eqref{activeuniformflux} and the circles depict the results from Brownian dynamics simulations. }
\label{uniformfig}
\end{figure}

In Fig.~\ref{uniformfig} (a-d) we show the density, stationary flux and the $x$ and $y$ components of the orientation, respectively. Note that despite the translational invariant motion in the $y$ direction there exists polarization in this direction. However, the $y$ component of the stationary flux is zero due to the cancellation of fluxes arising from the polarization in the $x$ and $y$ directions.

\section{Inhomogeneous magnetic field}
\label{spacedependent}

\begin{figure}
\centering
\resizebox*{1\linewidth}{5cm}{\includegraphics{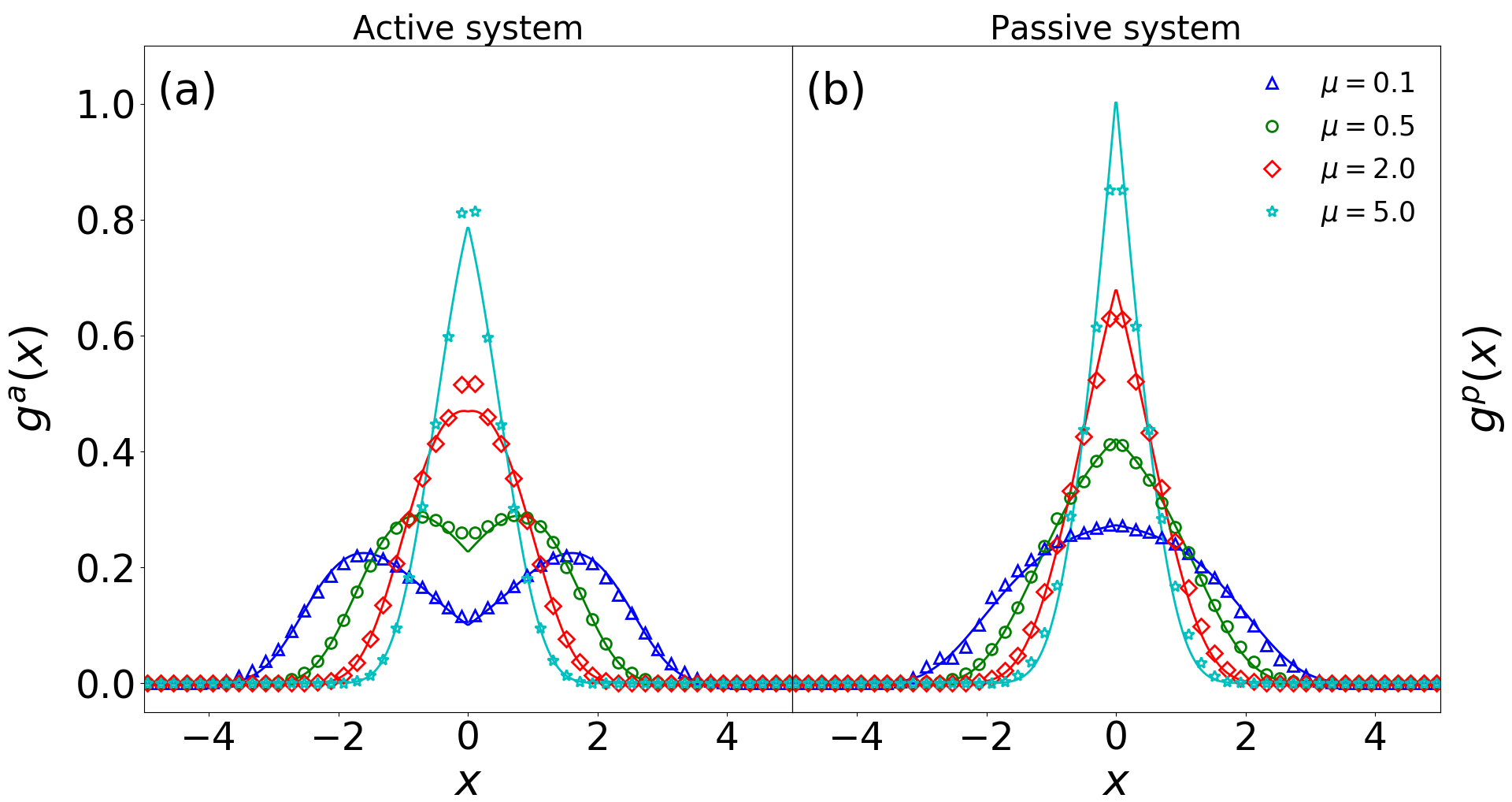}}
\caption{Probability density in (a) the active system and (b) the passive system for different values of $\mu$. The systems are subjected to a spatially inhomogeneous magnetic field such that $\kappa(x)=\sqrt{e^{\lambda|x|}-1}$ with $\lambda=2.0$. For the passive system, the translational diffusivity, $D_t$ has the same value as the active diffusivity, $D_a$. 
 While for the passive system the accumulation of particles is in a vicinity of  $x=0.0$, in the active system it is non-monotonic with local maxima at $x=\pm(2/\lambda)\ln\left(\lambda/2\alpha_a\right)$ for $\mu<\lambda^2v_0^2/(8D_r)$. The lines show the theoretical results from Eq.~\eqref{densityIA} and Eq.~\eqref{densityIP} and the symbols depict simulation results.}
\label{density}
\end{figure}

In this section, we show that novel features emerge in the case of an inhomogeneous magnetic field which have no counterpart in passive systems.  
We consider a system subjected to an exponentially varying magnetic field such that $\kappa(x)=\sqrt{e^{\lambda|x|}-1}$ where $\lambda$ is a constant. With this choice of the magnetic field, the Fokker-Planck equation in \eqref{fpeoned} can be solved exactly. The stationary probability density is given as
\begin{equation}
\label{densityIA}
g^a(x) = \frac{\alpha_a}{2L}\exp\left[\frac{\lambda|x|}{2}-\frac{2\alpha_a}{\lambda}\left(\exp(\frac{\lambda|x|}{2})-1\right)\right].
\end{equation}

Using Eq.~\eqref{polarization} the polarization in the $x$ direction is given as
\begin{subequations} \label{polarIA}
\begin{equation}
    p_x(x) = \frac{l_p\sign(x)}{2}\left[\frac{\lambda}{2}\exp\left(-\lambda|x|\right)+\alpha_a\exp(\frac{-\lambda|x|}{2})\right],
     \label{polarxIA}
\end{equation}
and similarly 

\begin{equation}
p_y(x) = \frac{l_p\lambda\sign(x)}{4\sqrt{\exp(\lambda|x|)-1}}\left[\frac{4\alpha_a}{\lambda}\sinh(\frac{\lambda|x|}{2})-\exp(-\lambda|x|)\right],
     \label{polaryIA}
\end{equation}
\end{subequations}
is the polarization in the $y$ direction.
The $x$ and $y$ components of the stationary flux can be obtained using Eq.~\eqref{fluxa1d}, which read

\begin{subequations} \label{fluxIA}

\begin{equation}
    j_x^a(x) = \alpha_aD_a\sign(x)\exp\left(-\frac{\lambda|x|}{2}\right)g^a(x),
     \label{fluxxIA}
\end{equation}
\begin{equation}
    j_y^a(x) = -\frac{\lambda D_a\exp\left(-\lambda|x|\right)}{2\sqrt{\exp(\lambda|x|)-1}}\sign(x)g^a(x).
     \label{fluxyIA}
\end{equation}
\end{subequations}

Note that the stationary polarization and stationary flux in the $y$ direction cease to exist in the absence of the magnetic field.

We also consider a passive system under resetting and subjected to the same magnetic field as in the active system. The governing Fokker-Planck equation  for the system can be easily written by setting $v_0=0$ in Eq.~\eqref{fluxa1d} and substituting Eq.~\eqref{fluxth1d} into Eq.~\eqref{fpeoned}. The stationary solution, $g^p(x)$, of the resulting equation is

\begin{equation}
\label{densityIP}
g^p(x) = \frac{\alpha_p}{2L\K_0(\frac{2\alpha_p}{\lambda})}\exp\left(\frac{\lambda|x|}{2}\right)\K_1\left(\frac{2\alpha_p}{\lambda}\exp\left(\frac{\lambda|x|}{2}\right)\right),
\end{equation}
where $\alpha_p=\sqrt{\mu/D_t}$ and $\K_0$ and $\K_1$ are the modified Bessel functions of the second kind of order 0 and 1, respectively. The $x$ component of the stationary flux can be written as
\begin{align}
\label{fluxxIP}
 j_x^p(x) = \frac{D\alpha_p^2}{2L\K_0(\frac{2\alpha_p}{\lambda})}\sign(x) \K_0\left(\frac{2\alpha_p}{\lambda}\exp\left(\frac{\lambda|x|}{2}\right)\right),
\end{align}
and similarly
\begin{equation}
    j_y^p(x) = -\sqrt{e^{\lambda|x|}-1} j_x^p(x).
     \label{fluxyIP}
\end{equation}
is the flux in the $y$ direction. 

\begin{figure}
\centering
\resizebox*{1\linewidth}{5cm}{\includegraphics{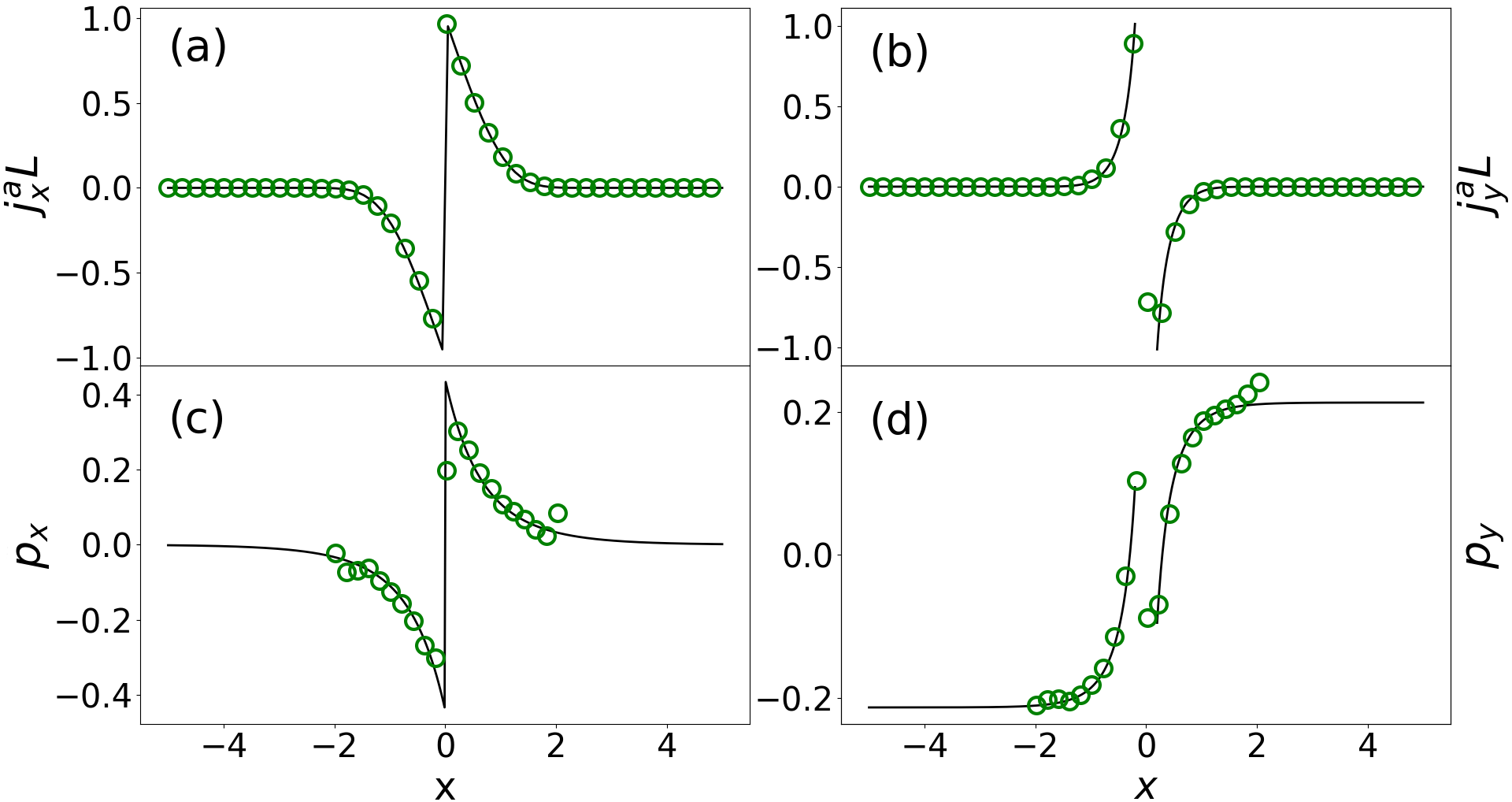}}
\caption{The $x$ and $y$ components of the flux and orientation are shown in (a) to (d), respectively. An active particle is stochastically reset to the line $x=0.0$ at the rate $\mu=2.0$. the particle is subjected to the magnetic field $\kappa(x)=\sqrt{e^{\lambda|x|}-1}$ with $\lambda=2.0$. The solid lines show the analytical solutions from Eqs.~\eqref{polarxIA} to \eqref{fluxyIA} and the circles depict the results from Brownian dynamics simulations. Note that the polarization and flux in the $y$ direction cease to exist in the absence of the magnetic field.}
\label{activeIB}
\end{figure}

Figure~\ref{density}(a) and Fig.~\ref{density}(b) show the probability density in the active and passive systems, respectively. While for the passive system the particles accumulate in a vicinity of $x=0$ for different values of the parameters, in the active system there exists an activity-dependent threshold rate such that for smaller resetting rates, the density distribution of the particles becomes non-monotonic. Below this threshold rate, $\mu<\lambda^2v_0^2/(8D_r)$, the ABPs accumulate in a vicinity of positions given by $x=\pm (2/\lambda)\ln\left(\lambda/2\alpha_a\right)$ .

\begin{figure}
\centering
\resizebox*{1\linewidth}{!}{\includegraphics{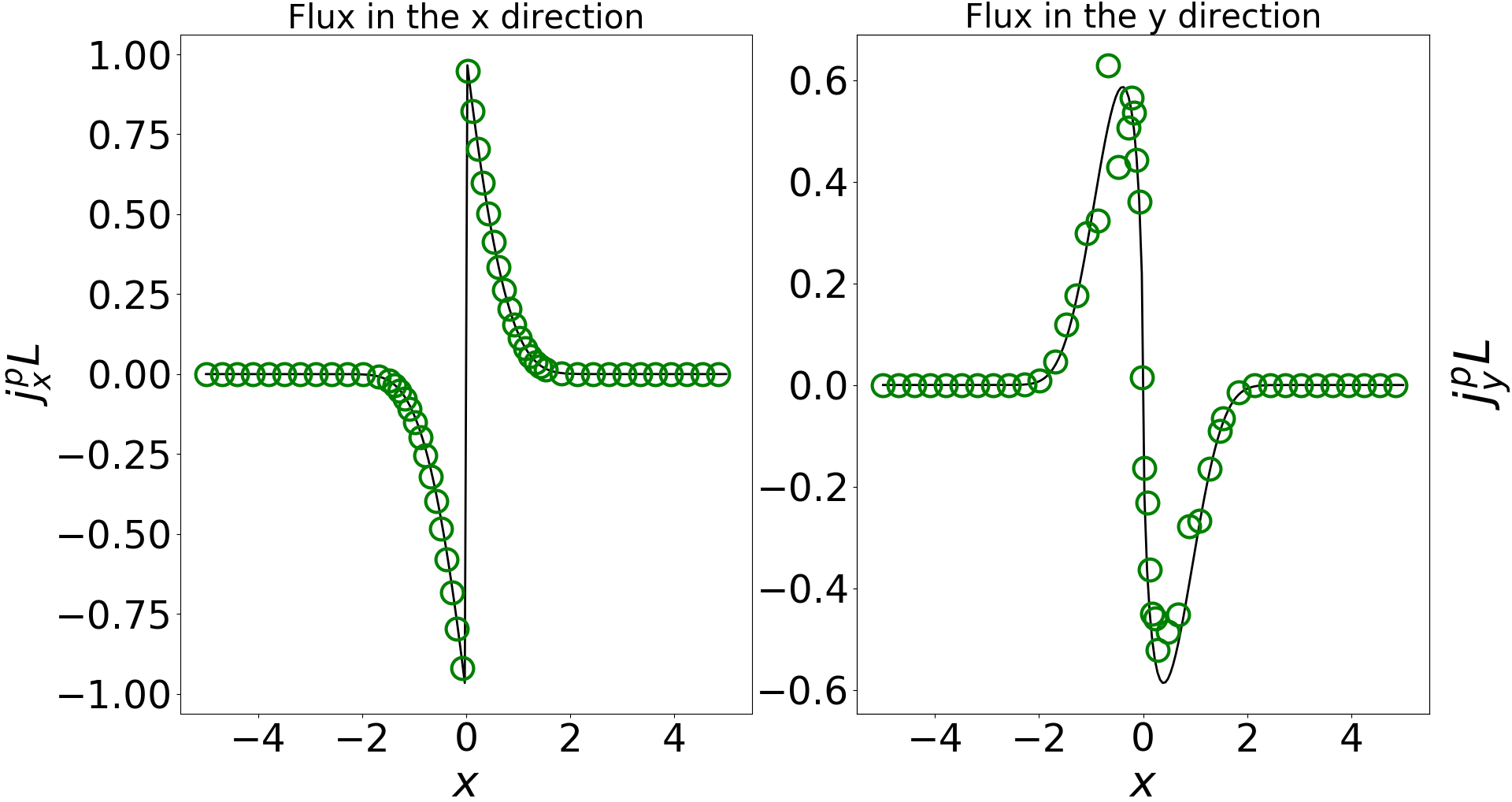}}
\caption{Flux in the $x$ and $y$ directions are shown in (a) and (b), respectively. A passive particle is stochastically reset to the line $x=0$ at the rate $\mu=2.0$. The particle is subjected to the magnetic field $\kappa(x)=\sqrt{e^{\lambda|x|}-1}$ with $\lambda=2.0$. The translational diffusivity, $D_t$ has the same value as the active diffusivity, $D_a$. The solid lines show the analytical solutions from Eqs.~\eqref{fluxxIP} and \eqref{fluxyIP} and the circles depict the results from Brownian dynamics simulations. }
\label{flux_passive}
\end{figure}

In Fig.~\ref{activeIB} we use Eq.~\eqref{polarxIA} to Eq.~\eqref{fluxyIA} to plot 
the fluxes and the polarization in the active system. Whereas in the case of a constant magnetic field there is no flux in the $y$ direction, inhomogeneity in the magnetic field gives rise to the orientation which results in fluxes in the $x$ and $y$ directions. Note that the polarization and flux in the $y$ direction cease to exist in the absence of the magnetic field. Figure~\ref{flux_passive} shows the $x$ and $y$ components of the stationary flux in the passive system. As can be seen the theoretical results from Eq.~\eqref{fluxxIP} and Eq.~\eqref{fluxyIP} are in good agreement with the simulation results. 

\begin{figure*}
\centering
\resizebox*{0.49\linewidth}{6.6cm}{\includegraphics{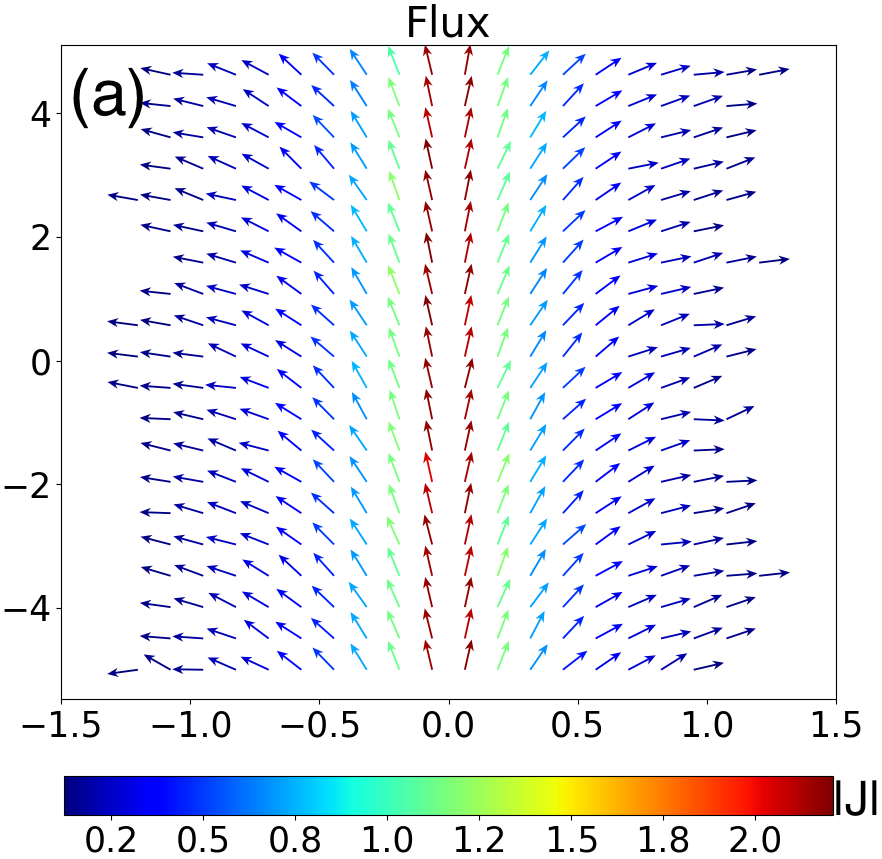}}
\resizebox*{0.5\linewidth}{6.5cm}{\includegraphics{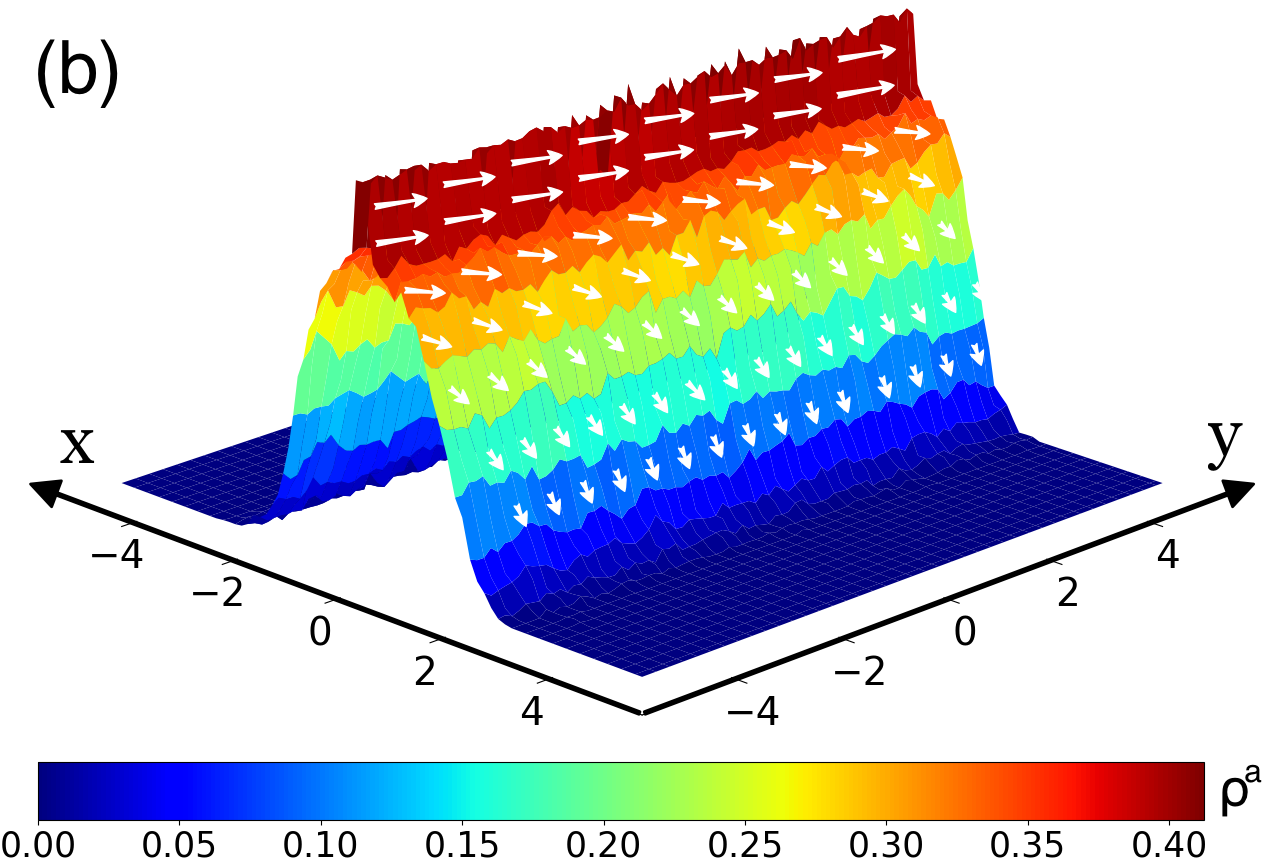}}
\caption{A spatial control of Lorentz force (or self-propulsion speed) can direct transport with no need for structured geometries. (a) A vector plot of the stationary flux whose direction is shown by the arrows and the magnitude is color coded and (b) a surface plot of the stationary probability density to which the flux is attached. An ABP under resetting to the line $x=0$ at the rate $\mu=1.0$ is subjected to the magnetic field such that $\kappa(x)=\sqrt{e^{-\lambda x}-1}$ if $x<0$ and $\kappa(x)=-\sqrt{e^{\lambda x}-1}$ otherwise with $\lambda=2.0$. }
\label{transport}
\end{figure*}
Transport properties of Brownian particles have been usually studied by considering systems which are restricted within the confines of structured and inhomogeneous environments. While in many cases, such structured environments can be viewed as confined channels with different boundaries and properties~\cite{malgaretti2019driving, ai2019collective,malgaretti2019special, li2020particle, bressloff2020modeling}, directed transport can be obtained via spatial control of activity~\cite{stenhammar2016light,sharma2017brownian}. Here we show that the Lorentz force can result in directed transport with no need for structured geometries. For a better visualization, we consider an ABP under resetting, subjected to the magnetic field $\kappa(x)=\sqrt{e^{-\lambda x}-1}$ if $x<0$ and $\kappa(x)=-\sqrt{e^{\lambda x}-1}$ otherwise. We show the flux and density in the active system in two dimensions. Figure~\ref{transport} (a) depicts a vector plot of the stationary flux in the system which clearly shows the particle transport along the $y$ axis. In Fig.~\ref{transport} (b) we show a surface plot of the stationary probability density in which the arrows show the direction of the particle transport.

\section{Mean first-passage time}
\label{MFPT_sec}

We now study the first-passage properties of the system in the case of a fixed target at the origin. The searching particle is stochastically reset to its initial position $x_0$ to be fixed. The backward Fokker-Planck equation for the survival probability, $G(x;t)$ -- the probability that the searching particle starting at $x$ at $t=0$ has not reached the target in time $t$ -- can be written as

\begin{align}
\label{backwardfpe}
\frac{\partial G(x;t)}{\partial t} & = A(x)\frac{\partial^2 G(x;t)}{\partial x^2} + B(x)\frac{\partial G(x;t)}{\partial x}  -\mu G(x;t)+\mu G(x_0;t),
\end{align}
where the initial and boundary conditions are $G(x;0)=1$ and $G(0;t)=0$, respectively. While the coefficients $A(x)$ and $B(x)$ for the active system are $D_ae^{-\lambda x}$ and $-D_a\lambda e^{-\lambda x}/2$, those for the passive one are $De^{-\lambda x}$ and $-D\lambda e^{-\lambda x}$, respectively. We first solve Eq.\eqref{backwardfpe} and then set $x$ to $x_0$ to find the MFPT (see Appendix~\ref{appendixB} for details). The Laplace transform of the backward Fokker-Planck equation in \eqref{backwardfpe} reads
\begin{align}
\label{laplacetransform}
B(x)\frac{\partial^2\tilde{G}(x;s)}{\partial x^2}  + A(x)\frac{\partial\tilde{G}(x;s)}{\partial x} &-(\mu+s)\tilde{G}(x;s)  =-1-\mu\tilde{G}(x_0;s),
\end{align}
where $\tilde{G}(x;s)=\int_0^\infty\dif t e^{-st}G(x;s)$ is the Laplace transform of the survival probability. Solving Eq.~\eqref{laplacetransform} and setting $x=x_0$ we obtain the expressions for the survival probability for the active and passive systems in the Laplace space, which when evaluated at $s=0$ gives the MFPT as
\begin{equation}
\label{mfptactive}
T^a(x_0)=\frac{1}{\mu}\left[\exp\left(\frac{2\alpha_a}{\lambda}\left(\exp(\frac{\lambda x_0}{2})-1\right)\right)-1\right],
\end{equation}
for the active system, and 
\begin{equation}
\label{mfptpassive}
T^p(x_0) = \frac{1}{\mu}\left[\frac{\K_1\left(\frac{2\alpha_p}{\lambda}\right)\exp(-\frac{\lambda x_0}{2})}{\K_1\left(\frac{2\alpha_p}{\lambda}\exp(\frac{\lambda x_0}{2})\right)}-1\right],
\end{equation}
for the passive system. Note that the MFPTs of the systems diverge as $\mu\rightarrow 0$ or $\mu\rightarrow\infty$. This implies that there exists an optimal rate at which the MFPT becomes minimum.

\begin{figure}[t]
\centering
\resizebox*{1\linewidth}{!}{\includegraphics{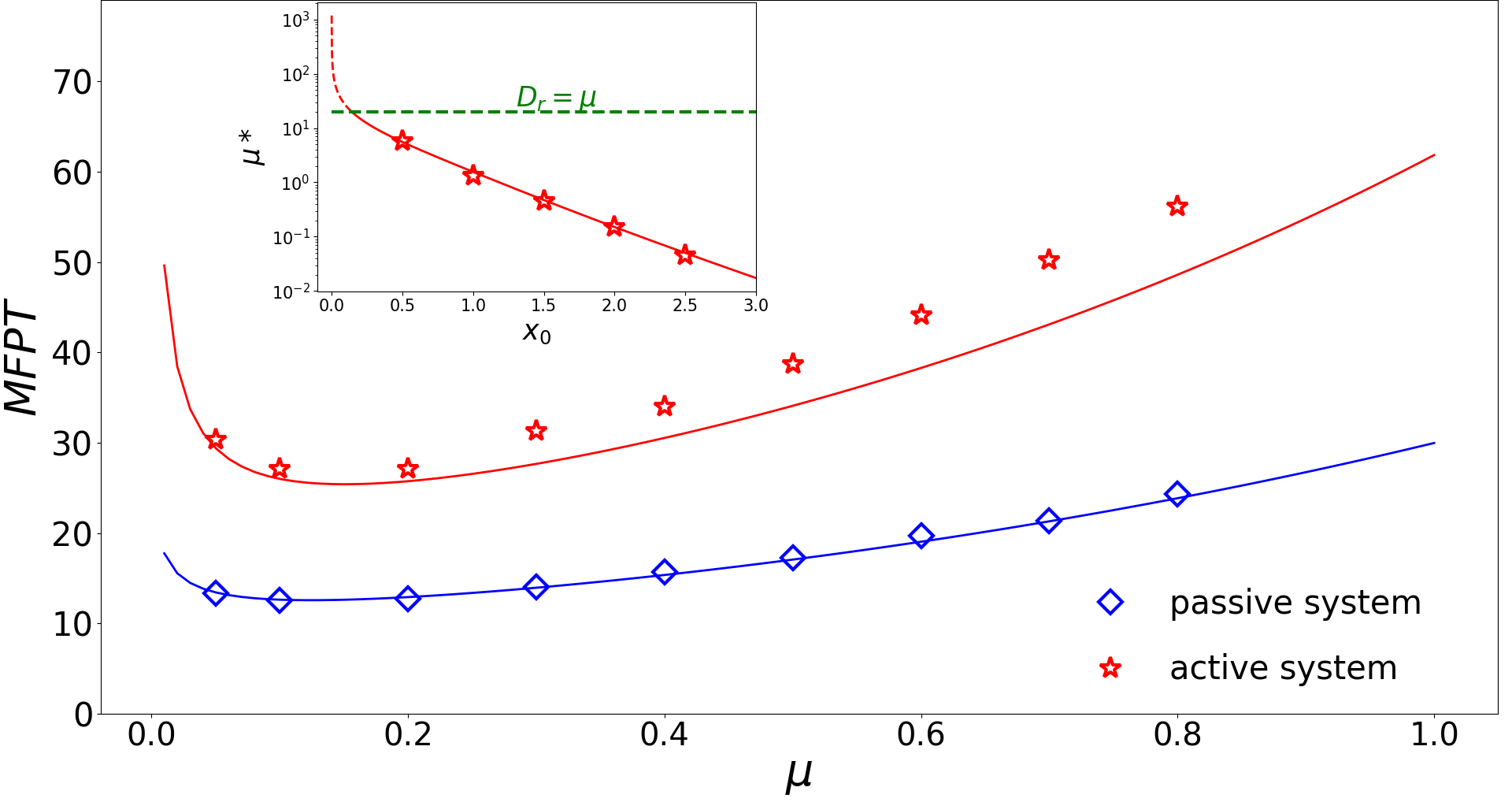}}
\caption{Mean first-passage time in the active and passive systems are shown in red and blue, respectively. The systems are subjected to the magnetic field such that $\kappa(x)=\sqrt{e^{\lambda|x|}-1}$ with $\lambda=2.0$. For the passive system, the translational diffusivity, $D_t$ has the same value as the active diffusivity, $D_a$. The solid lines show the theoretical predictions from Eq.~\eqref{mfptactive} and Eq.~\eqref{mfptpassive} and the symbols depict the results from Brownian dynamics simulations. The inset shows the optimal resetting rate with respect to the initial position $x_0$. The numerical solution of Eq.~\eqref{mfptactive} is compared with the simulation results.}
\label{MFPT}
\end{figure}

In Fig.~\ref{MFPT} we show the MFPT with respect to the stochastic rate for the active and passive systems. We compare the results from the theory, given by Eq.~\eqref{mfptactive} and Eq.~\eqref{mfptpassive} and those from Brownian dynamics simulations. It is clear that there exists an optimal resetting rate, $\mu^*$ that minimizes the time for the searcher to reach the target.  The optimal resetting rate decreases exponentially with increasing starting point $x_0$ due to inhomogeneity in the magnetic field. The inset shows how the optimal resetting rate varies with increasing initial position of the particle in the active system. 

\begin{figure}[t]
\centering
\resizebox*{1\linewidth}{!}{\includegraphics{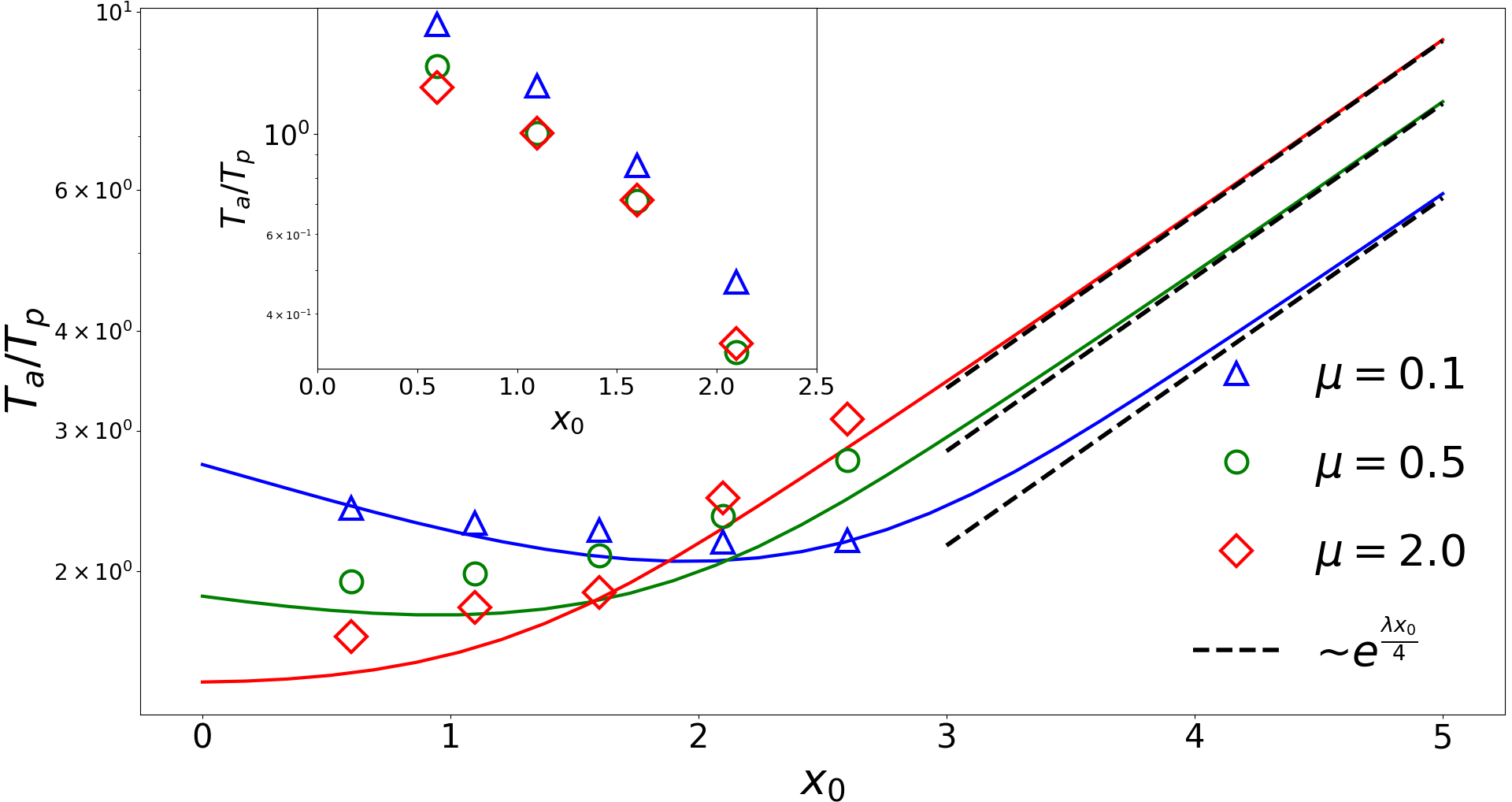}}
\caption{The ratio of the MFPT of the active particle to the passive passive one for different values of $\mu$. The systems are subjected to the magnetic field such that $\kappa(x)=\sqrt{e^{\lambda|x|}-1}$ with $\lambda=2.0$. For the passive system, the translational diffusivity, $D_t$ has the same value as the active diffusivity, $D_a$. The solid lines show the theoretical predictions from Eq.~\eqref{mfptactive} and Eq.~\eqref{mfptpassive} and the symbols depict the results from Brownian dynamics simulations. For a fixed target at the origin, the active particle is slower than the passive one. The relative slowness increases as $x_0\rightarrow 0$ or $x_0\rightarrow \infty$. It implies that there exists a position, $x_0^*$ where if the particles start from, they reach the target with minimum time difference. In the limit of large $x_0$ the MFPT for the active system to find the target is exponentially longer than the passive one and scales as $\sim e^{\lambda x_0/4}$,
which is shown by dashed line. The inset depicts the simulation results of the ratio of the MFPTs for a searcher whose initial position is the origin and the target is fixed at $x_0$. In this case, either active or passive searcher can be faster. There is also a point at which if the particles started from, they would have the same MFPT which occurs in the case of a constant magnetic field (e.g. $\lambda=0$) as well.}
\label{ratio}
\end{figure}

Figure~\ref{ratio} shows the ratio of the MFPT of the active system to its passive counterpart.
Interestingly, the active particle is slower than its passive counterpart to reach the target. The relative slowness increases as $x_0\rightarrow 0$ or $x_0\rightarrow \infty$. It implies that there exists a position, $x_0^*$ where if the particles start from, they reach the target with minimum  time difference. In the limit of large $x_0$ the MFPT for the active system to find the target is exponentially longer than the passive one and scales as $\sim e^{\lambda x_0/4}$,
which is shown by dashed line. The inset depicts the simulation results of the ratio of the MFPTs for the searcher starting at the origin and the target is set at $x_0$. In this case, either active or passive searcher can be faster. There is also a point at which if the particles started from, they would have the same MFPT which occurs in the case of a constant magnetic field (e.g. $\lambda=0$) as well.


\section{Conclusions}
\label{conclusion}

In this paper, we studied the motion of a charged ABP under resetting and the effect of Lorentz force. We showed that whereas for a uniform magnetic field the properties of the stationary state of the active system can be obtained from its passive counterpart, novel features emerge in the case of an inhomogeneous magnetic field which have no counterpart in passive systems. In particular, there exists an activity-dependent threshold rate such that for smaller resetting rates, the density distribution of active particles becomes non-monotonic. Moreover, somewhat counter intuitively, it may take an active particle much longer to reach a fixed target than its passive counterpart in an inhomogeneous magnetic field. We also showed that the Lorentz force can result in directed transport with no need for structured geometries.

We would like to emphasize that the choice of the magnetic field is motivated by the mathematical convenience, which allows us to theoretically analyse the system. The qualitative behaviour of the system will remain unaffected by other choices of the magnetic field. We note that an ABP in an inhomogeneous activity field and subjected to a constant magnetic field will give rise to the same phenomenology as presented in this study. A possible experimental realization is to reset the particle in a rotating frame of reference using optical tweezers. By rotating the reference frame one can induce a Coriolis force which acts the same as the Lorentz force arising from an external magnetic field~\cite{kahlert2012magnetizing}. From a future perspective, it would be interesting to investigate the effect of stochastic resetting on inertial ABPs~\cite{mandal2019motility,caprini2020inertial}.

\appendix 
\section{Elimination of orientational degrees of freedom}
\label{appendixA}
We use a gradient-expansion approach to integrate out the orientational degrees of freedom from the probability density. To do this, we recall the Fokker-Planck equation for the probability density, $P(\rr,\p;t)$ as   

\begin{align}
\label{fullfpeapp}
\frac{\partial}{\partial t} P(\rr, \p;t) &  =  \nabla\cdot\left[\Gama^{-1}(\rr)\cdot \left(D_t\nabla-v_0\p\right)P(\rr, \p;t)\right] \nonumber \\ 
 & + D_r \sR^2 P(\rr, \p;t)+\Phi_l + \Phi_g, 
\end{align} 
where 
\begin{equation}
\label{loss}
\Phi_l = - \mu P(\rr, \p;t),
\end{equation}
is the loss of the probability from the position $\rr$ due to resetting while
\begin{equation}
\label{gain}
\Phi_g = \mu\delta(x)\int P(x',y, p_x,p_y;t)\dif x',
\end{equation}
is the gain of the probability at the point $(0, y)$ on the $x$-axis.

The probability density, $P(\rr,\p;t)$, can be projected on spherical harmonics and consequently written as an expansion. By projecting the probability density on the zeroth and first spherical harmonics we find
\begin{equation}
\label{expansion}
P(\rr,\p;t) = \rho(\rr;t) + \sigg(\rr;t)\cdot\p + \Xi,
\end{equation}
where $\rho(\rr;t)$ is the positional probability density and the vector $\sigg(\rr;t)$ is the polar order parameter. Note that $\Xi$ denotes higher-order contributions which are ignored in this study. Plugging this expansion in Eq.~\eqref{fullfpeapp} and then integrating over the orientational degrees of freedom, we find
\begin{align}
\label{firststep}
\frac{\partial \rho(\rr;t)}{\partial t} & = -\nabla\cdot\left[\J(\rr;t)+\J^{a}(\rr;t)\right] + \phi_l + \phi_g,
\end{align} 
where
\begin{equation}
\label{loss}
\phi_l = - \mu \rho(\rr;t),
\end{equation}
\begin{equation}
\label{gain}
\phi_g = \mu\delta(x)\int \rho(x',y;t)\dif x',
\end{equation}
and 
\begin{subequations} \label{fluxes_app}
\begin{equation}
    \J(\rr;t) = - D_t\Gama^{-1}(\rr) \nabla \rho(\rr;t),
     \label{flux_app}
\end{equation}
\begin{equation}
    \J^{a}(\rr;t) = - \frac{v_0}{2}\Gama^{-1}(\rr) \sigg(\rr;t).
     \label{fluxa_app}
\end{equation}
\end{subequations}
To calculate the polar order parameter, $\sigg(\rr;t)$ we multiply Eq.~\eqref{fullfpeapp} by $\p$ and integrate over the orientational degrees of freedom. This gives rise to an equation for $\sigg(\rr;t)$ as
\begin{align}
\label{sigequation}
\frac{\partial}{\partial t}\sigg(\rr;t)&=\nabla\cdot\left[\Gama^{-1}(\rr)\left(D_t\nabla\sigg(\rr;t) - v_0\rho(\rr;t)\right)\right] \nonumber \\
& -(D_r + \mu)\sigg(\rr;t) +\mu\delta(x)\int\sigg(x',y;t)\dif x',
\end{align}

This is the point where we make an assumption in which the density is the slowest mode in the system. In this limit, the gradients in the system are small in comparison to the modified persistence length of the active particle, $l_p=v_0/(D_r+\mu)$ with $D_r\gg\mu$. With this approximation the time derivative of $\sigg(\rr;t)$ and the first term on the right hand side of Eq.~\eqref{sigequation}, sitting in the bracket, is negligible. In addition, due to the symmetry in the system the integral over the polar order, given as the last term in Eq.~\eqref{sigequation}, is zero. In this limit, Eq.~\eqref{sigequation} gives
\begin{equation}
\label{polar}
\sigg(\rr;t) = -\frac{v_0}{(D_r+\mu)}\nabla\cdot\left[\Gama^{-1}(\rr)\rho(\rr;t)\right].
\end{equation}

The substitution of Eq.~\eqref{polar} into Eq.~\eqref{fluxa_app} and then the resulting equation together with Eq.~\eqref{flux_app} into Eq.~\eqref{firststep} gives
\begin{align}
\label{fpe_app}
\frac{\partial \rho(\rr;t)}{\partial t}  = \nabla\cdot\left[\Gama^{-1}(\rr)\cdot\left(D_t\nabla+v_0\pii(\rr;t)\right)\rho(\rr;t)\right]+\phi_l + \phi_g,
\end{align}
where (in two dimensions) the polarization, $\pii(\rr;t)$ is related to the polar parameter, $\sigg(\rr;t)$ through
\begin{align}
\label{polarization_app}
\pii(\rr;t)  = \frac{\sigg(\rr;t)}{2\rho(\rr;t)}. 
\end{align}


\section{Derivation of the MFPT}
\label{appendixB}
Here we present the derivation of the MFPT of the active system. A similar approach can be used to derive the MFPT of the passive system. Using Eq.~\eqref{backwardfpe} in the main text, the backward Fokker-Planck equation for the active system can be written as 
\begin{align}
\label{backwardfpe_app}
\partial_t G(x;t) & = D_ae^{-\lambda x}\left[\frac{\partial}{\partial x^2}G(x;t)-\frac{\lambda}{2}\frac{\partial}{\partial x}G(x;t)\right]\nonumber \\ 
& -\mu G(x;t)+\mu G(x_0;t),
\end{align}
where $G(x_0;t)$ is the probability that the particle, started at $x_0$, is not absorbed by the target. The initial and boundary conditions are given as
\begin{equation}
  \begin{cases}
    G(x;t)=0, & \text{at $ t=0 $},\\
    G(x;t)=1, & \text{at $ x=0 $}.
  \end{cases}
  \label{conditions}
\end{equation}

Using the Laplace transform of $G(x;t)$, which is defined as 
\begin{equation}
\label{LT}
\tilde{G}(x;s)=\int_0^\infty e^{-st}G(x;s)\dif t,
\end{equation}  
the transformed backward equation can be written as
\begin{align}
\label{transformed_app}
D_ae^{-\lambda x} & \left[\frac{\partial}{\partial x^2}\tilde{G}(x;s) -\frac{\lambda}{2}\frac{\partial}{\partial x}\tilde{G}(x;s)\right]  -(\mu+s) \tilde{G}(x;s) \nonumber \\ 
&=-1-\mu \tilde{G}(x_0;s).
\end{align}

To solve Eq.~\eqref{transformed_app}, we first obtain the solution to the equation without the reinjection flux (the RHS terms), denoted by $\tilde{G}_0(x;s)$, i.e.
\begin{equation}
\label{homogeneousequation}
e^{-\lambda x}  \left[\frac{\partial}{\partial x^2}\tilde{G}_0(x;s) -\frac{\lambda}{2}\frac{\partial}{\partial x}\tilde{G}_0(x;s)\right]  -\alpha_a^2(s) \tilde{G}_0(x;s) = 0, 
\end{equation}
where $\alpha_a(s)=\sqrt{(\mu+s)/D_a}$. The solution to Eq.~\eqref{homogeneousequation} is
\begin{align}
\label{homogeneoussolution}
\tilde{G}_0(x;s) = A\exp\left[\frac{2\alpha_a(s)}{\lambda}\exp\left(\frac{\lambda x}{2}\right)\right] + B \exp\left[-\frac{2\alpha_a(s)}{\lambda}\exp\left(\frac{\lambda x}{2}\right)\right], 
\end{align}
where $A$ and $B$ are constants. The condition that the probability density is finite as $x\rightarrow\infty$ implies that $A=0$. Thus, the solution to the Eq.~\eqref{transformed_app} can be written as
\begin{equation}
\label{transformedsolution}
\tilde{G}(x;s) = B \exp\left[-\frac{2\alpha_a(s)}{\lambda}\exp\left(\frac{\lambda x}{2}\right)\right] + \frac{1+\mu\tilde{G}(x_0;s)}{\mu+s}, 
\end{equation}
where the constant $B$ can be calculated using the boundary condition in Eq.~\eqref{conditions}, which reads
\begin{equation}
\label{B}
B = -\frac{1+\mu\tilde{G}(x_0;s)}{\mu+s} \exp\left(\frac{2\alpha_a(s)}{\lambda}\right).
\end{equation}
Plugging Eq.~\eqref{B} into Eq.~\eqref{transformedsolution} gives 
\begin{align}
\label{finalsolution}
\tilde{G}(x;s)& = -\frac{1+\mu\tilde{G}(x_0;s)}{\mu+s}\exp\left[-\frac{2\alpha_a(s)}{\lambda}\left(\exp(\frac{\lambda x}{2}-1)\right)\right] \nonumber \\
& + \frac{1+\mu\tilde{G}(x_0;s)}{\mu+s}.
\end{align}
Finally, by setting $x=x_0$ we obtain the survival probability, which can be written as
\begin{equation}
\label{survivalprobability}
\tilde{G}(x_0;s)= \frac{1-\exp\left[-\frac{2\alpha_a(s)}{\lambda}\left(\exp(\frac{\lambda x_0}{2}-1)\right)\right]}{s+\exp\left[-\frac{2\alpha_a(s)}{\lambda}\left(\exp(\frac{\lambda x_0}{2}-1)\right)\right]},
\end{equation}
which when evaluated at $s=0$ gives the MFPT. This yields
\begin{equation}
\label{mfptactive_app}
T^a(x_0)=\frac{1}{\mu}\left[\exp\left(\frac{2\alpha_a}{\lambda}\left(\exp(\frac{\lambda x_0}{2})-1\right)\right)-1\right],
\end{equation}
where $\alpha_a\equiv\alpha_a(0)$. Note that we obtained the MFPT for the active system. However, a similar method can be used to derive the MFPT for the passive system, as well. 

\section{Acknowledgments}
A. Sharma acknowledges the support by the Deutsche Forschungsgemeinschaft (DPG) within the project SH 1275/3-1. 

\providecommand{\noopsort}[1]{}\providecommand{\singleletter}[1]{#1}%
\providecommand*{\mcitethebibliography}{\thebibliography}
\csname @ifundefined\endcsname{endmcitethebibliography}
{\let\endmcitethebibliography\endthebibliography}{}


\end{document}